# Modeling Loosely Annotated Images with “Imagined” Annotations

Hong Tang[1], Nozha Boujemma[2], Yunhao Chen[1]

1 ADREM, Beijing Normal University
2 IMEDIA project, INRIA, France

**Abstract**. In this paper, we present an approach to learning latent semantic analysis models from loosely annotated images for automatic image annotation and indexing. The given annotation in training images is loose due to: (1) ambiguous correspondences between visual features and annotated keywords; (2) incomplete lists of annotated keywords. The second reason motivates us to enrich the incomplete annotation in a simple way before learning topic models. In particular, some “imagined” keywords are poured into the incomplete annotation through measuring similarity between keywords. Then, both given and imagined annotations are used to learning probabilistic topic models for automatically annotating new images. We conduct experiments on a typical Corel dataset of images and loose annotations, and compare the proposed method with state-of-the-art discrete annotation methods (using a set of discrete blobs to represent an image). The proposed method improves word-driven probability Latent Semantic Analysis (PLSA-words) up to a comparable performance with the best discrete annotation method, while a merit of PLSA-words is still kept, i.e., a wider semantic range.

## 1 Introduction

Automatic image annotation is a process to use a computer program to assign keywords to images by learning from annotated images, where training images might be annotated by hand. Due to intensive labor and subjectivity of hand annotation, keywords used to be associated with images instead of specific regions, and only relevant to labelers. These images are referred to as loosely annotated images, for example Corel images used in [1]. The major challenge consists in (1) the ambiguous correspondence between visual features and annotated keywords; (2) the incompletion of annotated keywords. Some efforts have been devoted to learning from loosely annotated images, for instance learning latent semantic models [1-3], translating from discrete visual features to keywords [4-5], using cross-media relevance model [6-7], learning a statistic modeling for image annotation [8-11], image annotation using multiple-instance learning [12], and so on. In [13], the authors present a method to explicitly reduce the correspondence ambiguity between visual features and keywords before modeling the loosely annotated images. However, to our best knowledge, there is no effort explicitly dedicated to solving the incompletion of hand annotation before modeling. The possible reason might be that it is almost a same challenging task as automatically annotating new images. And it might be infeasible to solve a same difficult problem as an intermediate stage to achieve a final goal. We approach the incompletion of loosely annotated keywords in a simple way. In particular, we pick out and associate *missing* keywords in annotated training images with "imagined" occurrence frequencies by averaging similarity measures between them and annotated keywords. These retrieved missing keywords are referred to as "imagined" annotations. Then, words-driven probabilistic latent semantic analysis (PLSA-words [3]) is used to modeling both given and "imagined" annotations. At last, learned models are used to automatically annotate new images. Three example images and three kinds of annotations are illustrated in Fig. 1, where the second row corresponds to the "imagined" annotation of images.

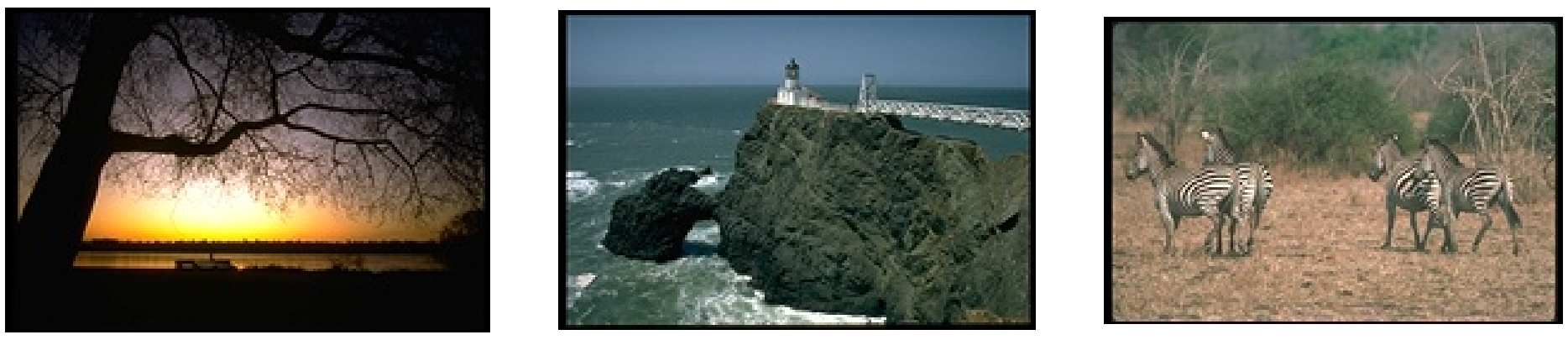

| | | |
|---|---|---|
| Bay Sun Tree | Coast Hill Water | Herd Plane Tree Zebra |
| *Sky Cloud Water Sea Forest* | *Wave Beach Boat Rock Sky* | *Jet Sky Grass Runway Elephant* |
| Sunset People Building Water Sky | Water Coast Ship Sky Mountain | Herd Zebra Elephant Water |

**Fig. 1** Three kinds of annotations of three example images. The first-row keyword is the loose annotation given in training images. The second row is the "imagined" annotation from all given annotations. The third row is automatic annotations when they are used as test images.

The rest of this paper is organized as follows. In section 2, we formulate the problem of enriching the incomplete annotation in the framework of automatic image annotation. The proposed algorithm to solving the problem is presented in section 3. Experimental results and discussions are given in section 4. Some conclusions are drawn in the last section.

## 2 The Problem

Given $N$ training images $D=\{I_1, I_2,\cdots, I_N\}$ with loose annotations, image $I_i$ is represented as a pair of histograms $(B_i,W_i)$. The first element is a blob-histogram with $p$ bins $B=\{b_1, b_2,\cdots, b_p\}$, where the bins are quantized from visual features of segmented regions. The second is a word-histogram with $q$ bins $W=\{w_1, w_2,\cdots, w_q\}$. $W_i^0$ and $W_i^+$ (resp. $B_i^0$ and $B_i^+$) denote two sets of bins whose values are equal to zero and non-zero, respectively. In other words, $W_i^0$ and $W_i^+$ are a set of non-annotated and annotated keywords in image $I_i$, respectively. Generally speaking, the number of annotated keywords, $|W_i^+|$, is far less than that of non-annotated keywords, $|W_i^0|$, for example, $|W_i^+|\leqslant 5$ and $|W_i^0|\approx 150$ in our experiments. Some non-annotated keywords should be used to describe semantics of image, but are *missing* occasionally. Our problem is to retrieve those *missing* keywords in a simple way before training images are used to learn a model for automatic annotation of new images. To discriminate them from annotated keywords given in training images, we refer to them as "imagined" annotations, since the process to obtaining them is associative and the results are not checked. To retrieve these missing keywords, assume a model $\theta_D$ has been learned from the training images $D$. We might predicate that a missing word $w_j \in W_i^0$ would be in the "imagined" annotations, if the conditional probability of the keyword given training image $I_i$

$$p(w_j \mid W_i^+, B_i, \theta_D) \geq \tau \,, \tag{1}$$

where $\tau$ ($0 \leqslant \tau \leqslant 1$) is a threshold to be determined in experiments. Another way is to pick out top $K$ keywords with higher conditional probabilities among all missing keywords $W_i^0$. As shown in Fig.1, top 5 keywords are picked out as "imagined" annotations for example images.

As we know, automatic image annotation is to assign some keywords to test image $I_{test}$ using a computer program. In general, a keyword $w_k$ might be assigned if it has a higher conditional probability among the word-vocabulary $W$

$$p(w_k \mid I_{test}, \theta) = p(w_k \mid B_{test}, \theta) \,, \tag{2}$$

where $B_{test}$ is the blob-histogram of test image $I_{test}$; $\theta$ is a learned model for automatic image annotation. Comparing the conditional probability in Eq. (1) and (2), we might conclude that retrieving missing keywords is also a process of automatic image annotation with an additional condition that missing keywords are dependent on annotated keywords $W_i^+$. Therefore, it might be at a same level of difficulty to learn the two models (i.e., $\theta_D$ and $\theta$), respectively. Since our final goal is to automatically annotate new images, we propose a simple algorithm to approximate the conditional probability in Eq. (1) instead of directly learning the $\theta_D$ from the training dataset $D$.

In the following, we outline two kinds of latent variable models for automatic image annotation, since they are closely related to the proposed algorithm. One is the probabilistic latent semantic analysis model [3, 14]. The other uses training image as latent variable, and annotate new images by summing out of all training images, for example [4, 5, 6].

If the learned model $\theta$ is the Probabilistic Latent Semantic Analysis model (PLSA) with $T$ topics, Eq. (2) becomes

$$p(w_k \mid I_{test}, \theta) = \sum_{t=1}^{T} p(t \mid I_{test}) p(w_k \mid t) \,, \tag{3}$$

where $p(t \mid I_{test})$ and $p(w_k \mid t)$ are model parameters to be estimated; the first parameter is a mixture coefficient of topics in the test image; the second is a distribution over keywords in the topic $t$. To estimate these parameters, one might maximize the log-likelihood of annotated keywords in $N$ training images $D$

$$\max_{\theta} L(D, \theta) = \max_{p(t \mid I_i); p(w_k \mid t)} \sum_{i=1}^{N} \sum_{k=1}^{q} w_{ik} \log \sum_{t=1}^{T} p(t \mid I_i) p(w_k \mid t) \,, \tag{4}$$

where $w_{ik}$ is the count of word $w_k$ in image $I_i$, i.e., the value of $k$th bin in word-histogram $W_i$. It can be seen that some terms in the log-likelihood $L(D, \theta)$ will not change with the estimated model parameters, when $w_{ik}$=0. In other word, missing keyword $w_k$ in the given annotation is out of consideration for estimating parameters in image $I_i$. Generally speaking, Eq. (4) is a reasonable criterion for parameter estimation, since maximizing likelihood is to generate all given observations in a joint probability as high as possible. In the word-histogram of training image $I_i$, $w_{ik}$=0 indicates that keyword $w_k$ was not observed in image $I_i$. Consequently, it is expected to be independent of the log-likelihood.

Due to the incompletion of loose annotation, even if $w_{ik}$=0, keyword $w_k$ might be still significantly relevant to content of the image. Assume we have pointed out this kind of keywords using estimated probability $p(w_k \mid W_i^+, B_i, \theta_D)$ in Eq. (1). Consequently, the likelihood in Eq. (4) would change with these keywords, and a "better" model would be expected. In the proposed algorithm, during learning topic models, the zero, i.e., $w_{ik}$=0, would be directly replaced with the conditional probability in Eq. (1) if it is larger than the given threshold $\tau$.

The other kind of latent variable models for automatic image annotation is to learn or approximate the conditional probability in Eq. (2) through summing out of all training images [11]

$$\begin{aligned} p(w_k \mid B_i, \theta) &\sim \sum_{i=1}^{N} p(w_k, B_i \mid I_i) \\ &= \sum_{i=1}^{N} p(w_k \mid I_i) p(B_i \mid I_i) \end{aligned}, \quad (5)$$

where the keyword $w_k$ and visual features $B_i$ are assumed to be conditional independence given image $I_i$. In this kind of methods, an uninformative prior is actually assumed for each training image. Consequently, the prior of keywords (or blobs) used to be unequal and is dependent on the given training dataset. In general, this kind of methods seem more straightforward and "simpler" to estimation than probabilistic topic models. In some cases, they are more effective than PLSA because of the "simplicity". However, they are biased to annotate images with frequent keywords (as shown in the later experiments). For the sake of "simplicity", we use a similar method to approximate the conditional probability in Eq. (1) in the proposed algorithm. Some missing keywords are picked out as "imagined" annotations in a simple way. Then, both given and imagined annotations are used to estimate parameters of PLSA models. In this sense, the propose algorithm is a combination of two kinds of latent variable models mentioned above. Accordingly, one might expect that the proposed algorithm would benefit from both kinds of latent variable models.

## 3 The Proposed Algorithm

The proposed algorithm includes two stages: (1) obtaining "imagined" annotations through approximating conditional probability of missing keywords given training images and loose annotations; (2) modeling both given and imagined annotations using PLSA-words [3]. For convenience of expression, we refer to the proposed algorithm as Virtual-Word driven Probabilistic Latent Semantic Analysis (PLSA-vw). The two stages of PLSA-vw are detailed in the following two subsections, respectively.

### 3.1 Obtaining "Imagined" Annotations

Before the process is specified, we describe how the conditional probability in Eq.(1) will be approximated. Assume that the conditional probability of a missing keyword $w_j$ is the average of all joint probabilities between it and keywords given in the loose annotation

$$p(w_j \mid W_i^+, B_i, \theta_D) \sim \frac{1}{|W_i^+|} \sum_{w_k \in W_i^+} p(w_j, w_k \mid D), \quad (6)$$

where $|W_i^+|$ is the number of annotated keywords in training image $I_i$. Please note that $w_j \in$

$W_i^0$ and $w_k \in W_i^+$; Actually, the approximation in Eq. (6) is achieved by assuming an uninformative prior for the join probability $p(w_j, w_k | D)$, which is furthermore approximated by summing out of all training images

$$\begin{aligned} p(w_j, w_k \mid D) &\sim \sum_{i=1}^{N} p(w_j, w_k \mid I_i) \\ &\sim \sum_{i=1}^{N} p(w_j, I_i) p(w_k, I_i) \end{aligned}, \tag{7}$$

where keywords $w_j$ and $w_k$ are assumed to be conditional independence given training images. At last, the joint probability that a keyword co-occurs with an image is approximated by

$$p(w_j, I_i) = \frac{w_{ij}}{\sum_{i=1}^{N} w_{ij}}, \tag{8}$$

where $w_{ij}$ is the count of keyword $w_j$ in image $I_i$. It is easy to see that the joint probability $p(w_j, w_k | D)$ in Eq. (7) is actually approximated by an inner product between two normalized word-count vectors or a cosine-like similarity measure between keywords.

Let count-matrix **W** (resp. **B**) be a set of word- (resp. blob-) histograms where each row corresponds to an training image. The actual approximation method is an inverse process from Eq. (6) to (8), as shown in the following four steps:

*Step 1*: Compute normalized word-count matrix $\mathbf{W}_{norm} = \frac{\mathbf{W}.}{\vec{e}_N \vec{e}_N^T \mathbf{W}}$.

*Step 2*: Compute similarity matrix between keywords $\mathbf{W}_{sim} = \mathbf{W}_{norm}^T \mathbf{W}_{norm}$.

*Step 3*: Approximate conditional probability matrix of keywords $\mathbf{W}_{img} = \frac{(\mathbf{W}\mathbf{W}_{sim}).}{\mathbf{W}\vec{e}_q \vec{e}_q^T}$.

*Step 4*: Pick out all missing keywords: (1) if $\mathbf{W}(i, j)>0$, $\mathbf{W}_{img}(i, j)$ is replaced with zero, since $w_j$ has been annotated in image $I_i$; (2) if $\mathbf{W}_{img}(i, j) \geqslant \tau$, $w_j$ is a missing keyword in image $I_i$, otherwise $\mathbf{W}_{img}(i, j)$ is set into zero.

In the above-mentioned steps, $\vec{e}_x$ is an $x$ dimensional column vector whose elements are equal to 1, and the matrix division is performed at every correspondent entry. Up to now, all non-zero entries in the matrix $\mathbf{W}_{img}$ correspond to keywords missed in the loose annotation, which are assumed relevant to semantics of images. All keywords retrieved from training images are referred to as *"imagined" annotations*, in contrast with the given annotations in training images.

### 3.2 Modeling both Given and Imagined Annotations

To automatically annotate new images using conditional probability of keywords given its visual representation, one might learn the model $\theta$ in Eq. (2) by maximizing log-likelihood as shown in Eq. (4). In our case, we have two kinds of observations, i.e., given and imagined annotations. Generally speaking, the imagined annotations would be not as reliable as given annotations. For example, as shown in Fig.1, the imagined annotation of the third example image includes " *Jet Sky Grass Runway Elephant*", where the "*Jet*" is obviously irrelevant to the image. As mentioned before, the imagined annotations are associative and are not checked by human supervision. Therefore, we simply regard the approximated conditional probability of missing keywords as their reliability in the imagined annotations. Furthermore, we use the approximated conditional probability as a *real-value* word-count. Typically, the real-value word-count is less than one. Therefore, we can define a new word-count matrix for learning

$$\mathbf{W}^{*} = \mathbf{W} + \mathbf{W}_{img}\,, \tag{9}$$

where $\mathbf{W}$ and $\mathbf{W}_{img}$ are word-count matrixes in given and imagined annotations, respectively. It can be seen from the process of approximation in subsection 3.1 that imagined annotations $\mathbf{W}_{img}$ are obtained bypassing blob-count matrix $\mathbf{B}$. In other words, these annotations have been imagined without consulting visual features of training images. To ensure the imagination can be reflected on visual features, we derive a new blob-count matrix for learning in the same way

$$\mathbf{B}^{*} = \mathbf{B} + \mathbf{B}_{img}\,, \tag{10}$$

where the imagined blob-count matrix $\mathbf{B}_{img}$ is obtained from matrix $\mathbf{B}$ using the same approximation method in subsection 3.1. In the process, the changes what we need make include (1) replacing matrixes $\mathbf{B}$ and $\mathbf{B}_{img}$ with $\mathbf{W}$ and $\mathbf{W}_{img}$, respectively; (2) accordingly, normalized word-count matrix $\mathbf{W}_{norm}$ and similarity matrix $\mathbf{W}_{sim}$ would be rewritten as $\mathbf{B}_{norm}$ and $\mathbf{B}_{sim}$; (3) the number of keyword, $q$, should be replaced with that of blob, $p$, in step 3. Then, PLSA-words is used to modeling new observation matrixes, i.e., $\mathbf{W}^{*}$ and $\mathbf{B}^{*}$. We outline PLSA-words in the following as two stages: (1) estimating parameters of probabilistic latent semantic analysis models; (2) automatic image annotation of a set of new images.

*Parameter estimation*: Unlike Eq. (4), the likelihood is given by

$$\max_{\theta} L(D,\theta) = \max \sum_{i=1}^{N}\sum_{j=1}^{q} \mathbf{W}^{*}(i,j) \sum_{t=1}^{T} \log p(t \mid I_i)\, p(w_j \mid t)\,, \tag{11}$$

where $\mathbf{W}^{*}(i, j)$ is the count of keyword $w_j$ in training image $I_i$ with both given and "imagined" annotations. By maximizing Eq. (11), both topic models of keywords, i.e., $p(w_j \mid t)$, $w_j \in W$, and mixture coefficients, i.e., $p(t \mid I_i)$, $t \in [1,T]$ and $I_i \in D$, in each training image can be estimated. Keeping $p(t \mid I_i)$ untouched, maximizing

$$\max \sum_{i=1}^{N}\sum_{j=1}^{p} \mathbf{B}^{*}(i,j) \sum_{t=1}^{T} \log p(t \mid I_i)\, p(b_j \mid t) \tag{12}$$

is to obtain topic models of blobs, i.e., $p(b_j \mid t)$, $b_j \in B$, where matrix $\mathbf{B}^{*}$ is given in Eq. (10). This process is referred to as folding-in visual features in [3].

*Automatic image annotation*: Given blob histograms of a set of test images $\mathbf{B}^{test}$, a new matrix $\mathbf{B}^{*test}$ is derived using $\mathbf{B}^{*test} = \mathbf{B}^{test} + \mathbf{B}^{test}_{img}$, where the imagined blob-count matrix of test images $\mathbf{B}^{test}_{img}$ is given by

$$\mathbf{B}^{test}_{img} = \frac{(\mathbf{B}^{test}\mathbf{B}_{sim}).}{\mathbf{B}^{test}\vec{e}_p \vec{e}_p^{\,T}}\,, \tag{13}$$

where the blob-similarity matrix $\mathbf{B}_{sim}$ has been derived from training images instead of test images.

Then, keeping learned topic models of blobs $p(b_j \mid t)$ , $t \in [1,T]$, untouched, one can obtain mixture coefficients $p(t \mid I_{test})$ in test image $I_{test}$ by maximizing Eq. (12), where $\mathbf{B}^*$ should be replaced with $\mathbf{B}^{*test}$. Then, using learned mixture coefficients $p(t \mid I_{test})$ in test image, one can compute the conditional probability of each keyword given any test image

$$p(w_j \mid B_{test}, \theta) = \sum_{t=1}^{T} p(t \mid I_{test}) p(w_j \mid t) , \tag{14}$$

where topic models of keywords, $p(w_j \mid t)$, $w_j \in W$, have been estimated in the stage of parameter estimation as shown in Eq. (11).

# 4 Experiments and Discussion

In this section, we evaluate the performance of the proposed algorithm from two aspects. First, we examine the relative improvement of PLSA-vw over PLSA-words in terms of image annotation, indexing and retrieval. Second, the proposed method is compared with three typical discrete annotation methods, i.e., machine translation (MT) [4], translation table (TT) [5], and cross-media relevance model (CMRM) [6]. Unlike PLSA-vw and PLSA-words, these methods use image as latent variable, and sum out of all training images to annotate new images [11]. Therefore, the annotation performance of these methods is heavily dominated by the empirical distribution of keywords in training images. As shown in subsection 4.2, these methods are biased to annotate images with frequent keywords.

## 4.1 Experimental Setting

We conducted experiments over a publicly available Corel dataset (http://kobus.ca/research/data/jmlr_2003), which was provided by Bernard et al. [1]. This dataset is organized into ten samples from about 16 000 annotated images, in which there are segmentation, hand annotation, visual features and its blob representation. Each sample is split into a training set and two test sets. The files of the three subsets correspond to no-prefix (training), "test_1_" and "test_3_" (novel), respectively. We use the first test set (i.e., "test_1_") as new images to be annotated. The average sizes of word and blob vocabulary are 161 and 500, respectively. Moreover, the translation tables used in the machine translation method [4] are also provided in the dataset. The number of topics used in our experiments is 120, and the threshold $\tau$ in Eq. (1) is set to 0.01, which was determined through cross-validation with 10% holdout training images.

We use four indexes (i.e., AP, mAP, RP and RSI) to evaluate the performance of algorithms, which are related to image annotation, indexing, retrieval and semantic range, respectively. Given a test image, top 5 keywords with higher conditional probability in Eq. (2) will be used as predicated annotations. The first index is the average Annotation Precision of all test images (AP), where the annotation precision of a test image is a ratio of the number of correctly annotated keywords to that of keywords in the ground-truth annotation. The second index is mean Average Precision (mAP), which is the mean value of average retrieval precisions of all keywords. The average retrieval precision of a keyword is defined as the sum of the retrieval precisions of the correctly retrieved words at its rank, divided by the total number of relevant images. Here, an image is termed as relevant if the querying keyword is in the ground-truth annotation. Therefore, mAP is sensitive to the entire ranking of image indexing by the conditional probability in Eq. (2). The third index is insensitive to the entire ranking of image indexing, i.e., average Retrieval Precision (RP) of using every keyword as querying. Given top *M* retrieved images, the retrieval precision is the number of images with the querying keyword divided by *M*. In our experiment, *M* is equal to 20. The last index is very similar to the "Range of Semantics Identified" (RSI), which was originally coded by Bernard et al. in [13, 15]. RSI is to measure how many keywords can be correctly identified during image annotation in [13]. Since the number of keywords in different samples is variable in our experiments, RSI is defined as what percentage of keywords is ever correctly annotated at least once. All indexes mentioned above are averaged over ten samples.

### 4.2 Results

Table 1 lists the performance of automatic annotation methods used in our experiments, where the number in brackets is the variance of ten samples. Given an index (a column), the best and next methods are marked with red and blue color, respectively. Although PLSA-vw is not the best for any index and only is the second for three of four indexes, it does, as expected, benefit from two kinds of latent variable models outlined in section 2.

**Table 1** Performance of automatic annotation methods in terms of four indexes (%). The number in brackets is the variance of 10 samples. The best and next performances in each column are marked as red and blue color, respectively.

| | AP | mAP | RP | RSI |
|---|---|---|---|---|
| *PLSA-vw* | 17.56 (1.34) | ***12.34*** (1.72) | ***15.89*** (3.59) | ***68.43*** (1.96) |
| *PLSA-words* | 14.06 (1.85) | 10.62 (1.98) | 13.82 (2.68) | **69.51** (2.35) |
| *TT* | **23.38** (1.36) | **12.49** (1.29) | **16.34** (2.16) | 33.48 (1.72) |
| *MT* | ***21.40*** (1.19) | 9.35 (0.94) | 12.76 (1.34) | 43.59 (2.35) |
| *CMRM* | 18.42 (0.98) | 10.89 (1.34) | 13.54 (1.50) | 13.08 (0.88) |

First of all, it can be seen that the performance of PLSA-vw is improved over PLSA-words in terms of image annotation (i.e., AP), indexing (i.e., mAP) and retrieval (i.e., RP). As shown in section 3, PLSA-words, in the proposed algorithm, is utilized to model both given and imagined annotations. Therefore, PLSA-vw only differs from PLSA-words in the input, i.e., imagined annotations, for modeling. It can be concluded that the improvements of PLSA-vw over PLSA-words do originate from the imagined annotations. To some extent, experimental results show that it is useful and might be necessary to enrich the incomplete list of loosely annotated keywords for automatically annotating new images. This is the focus of this paper. At the same time, we can see that the improvements are at a little cost of the semantic range, i.e., 1% RSI in table 1, although the cost is not significant relative to other methods. In the following, we compare the proposed algorithm with other three discrete annotation methods.

Although TT is better than PLSA-vw in terms of three of four indexes (i.e., AP, mAP and RP), PLSA-vw is comparable to TT for mAP and RP, and both of them are better than other methods. In terms of AP, the difference between them is significant and up to 5%. However, the annotation performance of TT would be over-estimated if we do not care about the other index, i.e., RSI. As shown in table 1, RSI of TT is equal to 33.48%. That means only 33.48% keywords are ever correctly annotated using TT. In other words, over 66% keywords are out of consideration in automatic annotation. Therefore, TT is heavily biased to using certain keywords as annotation of images. To show what kind of keywords is liable to be used in TT, frequencies of keywords in the first sample are shown in Fig. 2 (a), where red cross and blue dot respectively denote whether keywords are ever correctly annotated in all test images or not. It can be seen from Fig. 2 (a) that TT can correctly annotate only 2 keywords among keywords whose word-counts are less than 50. In contrast, PLSA-vw can correctly annotate more than 20 keywords as shown in Fig. 2 (b). For the same index, i.e., AP, PLSA-vw can outperform PLSA-words up to 3.5%. However, the cost of RSI is about 1%. In this sense, the improvement of PLSA-vw over PLSA-word seems more attractive, although PLSA-vw is still not the best one.

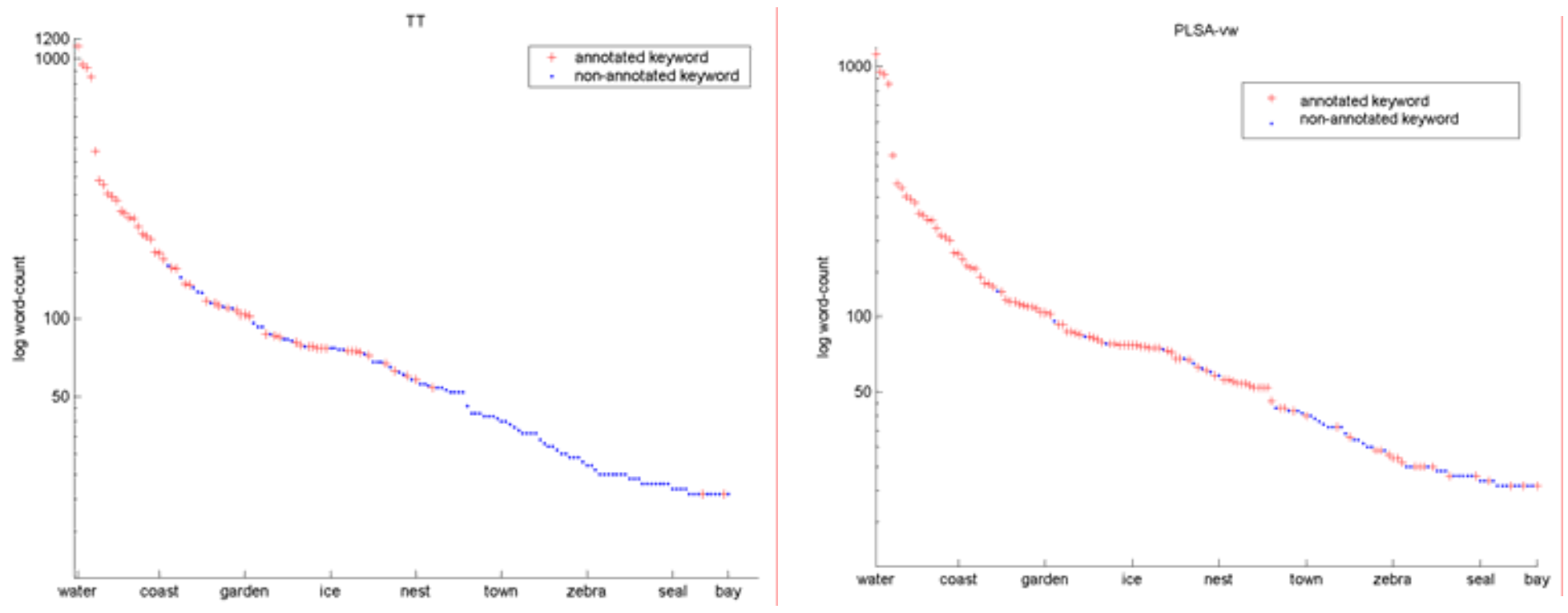

(a) TT (b) PLSA-vw

**Fig. 2** Word-count of all keywords in the first sample. Some keywords are displayed along horizontal axis. All annotated and non-annotated keywords are marked as red cross and blue dot, respectively. There are only two keywords are ever correctly annotated using TT when the word-count is less than 50. In the same time, PLSA-vw can correctly annotate more than 20 keywords.

## 5 Conclusions

In this paper, we present an approach to enriching the incomplete list of keywords. In particular, some missing keywords are "imagined" as real annotations for training images by approximating the conditional probability of missing keywords given loosely annotated images. Experimental results show that the proposed algorithm may improve the performance of PLSA-words in terms of image annotation, indexing and retrieval while keeping the merit of PLSA-words than other discrete annotation methods, i.e., wider semantic range. Therefore, it can be concluded that it is useful and necessary to enrich the incomplete list of keywords given in loosely annotated images. Although the proposed algorithm is rather simple, we make a very strong assumption, i.e., the process of retrieving missing keywords is conditional independent of visual features given a same process of imagination for blobs. Our furthermore work is to remove the assumption, and make the imagination of keywords depend directly on visual features of images.